\documentclass[reprint,prl,twocolumn,nofootinbib,aps]{revtex4-2}
\usepackage{amsmath}
\usepackage{amsfonts}
\usepackage{amssymb}
\usepackage{mathtools}
\usepackage{graphicx}
\usepackage[dvipsnames]{xcolor}
\usepackage[colorlinks=True,citecolor=myRed,linkcolor=myRed,urlcolor=gray]{hyperref}
\usepackage{hypcap}
\usepackage{braket}
\usepackage{yfonts}
\usepackage{bm}
\usepackage{ulem}
\usepackage{dsfont}

\newcommand\eq[1]{\begin{align}#1\end{align}}

\newcommand\Ct{\mathcal{C}}

\newcommand\Rt{\mathcal{R}}

\newcommand\nh{N_\mathcal{H}}

\newcommand\new[1]{{\color{black}{#1}}}

\definecolor{myBlue}{RGB}{31,119,180}
\definecolor{myOrange}{RGB}{255,127,14}
\definecolor{myGreen}{RGB}{44,160,44}
\definecolor{myRed}{RGB}{214,39,40}
\definecolor{myPurple}{RGB}{148,103,189}

\makeatletter
\def\p@figure{\color{myBlue}}
\def\p@equation{\color{myRed}}
\makeatother

\begin{document}

\title{The connection between Hilbert-space return probability and real-space autocorrelations in quantum spin chains}

\author{Bikram Pain}%$^\star$}
\email{bikram.pain@icts.res.in}
\affiliation{International Centre for Theoretical Sciences, Tata Institute of Fundamental Research, Bengaluru 560089, India}

\author{Kritika Khanwal}%$^\star$}
\email{kritika.khanwal@icts.res.in}
\affiliation{International Centre for Theoretical Sciences, Tata Institute of Fundamental Research, Bengaluru 560089, India}

\author{Sthitadhi Roy}
\email{sthitadhi.roy@icts.res.in}
\affiliation{International Centre for Theoretical Sciences, Tata Institute of Fundamental Research, Bengaluru 560089, India}

\begin{abstract}
The dynamics of interacting quantum many-body systems has two seemingly disparate but fundamental facets. The first is the dynamics of real-space local observables, and if and how they thermalise. The second is to interpret the dynamics of the many-body state as that of a fictitious particle on the underlying Hilbert-space graph. In this work, we derive an explicit relation between these two aspects of the dynamics. We show that the temporal decay of the autocorrelation in a disordered quantum spin chain is explicitly encoded in how the return probability on Hilbert space approaches its late-time saturation. As such, the latter has the same functional form in time as the decay of autocorrelations but with renormalised parameters. Our analytical treatment is rooted in an understanding of the morphology of the time-evolving state on the Hilbert-space graph, and corroborated by exact numerical results.
\end{abstract}

\maketitle

The dynamical phase diagram of interacting, often disordered, quantum many-body systems has been a topic of intense interest and scrutiny over the last several years. 
Fundamental questions in this context have revolved around the many-body localised (MBL) phase at strong disorder, the accompanying phase transition from an ergodic to the MBL phase (see Refs.~\cite{nandkishore2015many,abanin2017recent,alet2018many,abanin2019colloquium} for reviews and references therein), as well as the anomalously slow dynamical regime in the vicinity of the transition~\cite{lev2015absence,agarwal2015anomalous,luitz2016anomalous,gopalakrishnan2016griffiths,luitz2016extended,khait2016spin,znidaric2016diffusive,agarwal2017rare,roy2018anomalous,lezama2019apparent}. 
There are two complementary facets to this problem. 
The first is the dynamics of real-space correlation functions such as local autocorrelations or spatiotemporal correlations describing transport and entanglement growth.
The second is interpreting the dynamics in terms of that of a fictitious particle, a proxy for the many-body state, on the complex, correlated Hilbert-space graph.

Besides the understanding that the phenomenology of the MBL phase can be described in terms an extensive number of (quasi)local conserved quantities ~\cite{serbyn2013local,huse2014phenomenology,ros2015integrals,imbrie2017local}, an essential insight from the real-space picture is that systems near the MBL transition can be viewed as a patchwork of locally ergodic and locally MBL regions. 
As such, the anomalously slow dynamics in the ergodic phase preceding the transition is dominated by {\it rare Griffiths regions} where the disorder fluctuations are anomalously strong~\cite{agarwal2015anomalous,gopalakrishnan2016griffiths,agarwal2017rare,schiro2020toy,taylor2021subdiffusion,turkeshi2022destruction}. 
However, these theories are primarily phenomenological in nature. 
On the other hand, approaches based on the Hilbert-space graph is arguably more fundamental and deeper rooted in microscopics as the temporal evolution of the state amplitudes on the Hilbert space constitute the most basic ingredients from which almost every other dynamical correlation can be reconstructed~\cite{detomasi2020rare,roy2021fockspace,roy2022hilbert,creed2023probability,roy2023diagnostics}.
It is therefore of immanent importance to understand concretely the relations between simple dynamical quantities on the Hilbert space and spatiotemporal correlations in real space.
That being said, it is an extremely challenging task {\new {as the locality of the system in real space is encoded in a very intricate manner on the Hilbert-space topology\footnote{For instance, dynamics of a single spin corresponds to that of the fictitious particle along a specific axis of the Hilbert-space graph.}}}, and efforts in this direction are fledgling~\cite{detomasi2020rare,roy2021fockspace,roy2022hilbert,creed2023probability,roy2023diagnostics}. 

In this work, we take a step in this direction by showing that the return probability on the Hilbert space, possibly the simplest dynamical correlation therein~\cite{torres-herrera2015dynamics,torres-herrera2018generic,schiulaz2019thouless}, is simply and directly related to the local autocorrelation in disordered quantum spin chains. 
This is with the motivation that lessons from Anderson localisation on high-dimensional graphs~\cite{anderson1958absence,abou-chacra1973self,tikhonov2016anderson,biroli2017delocalized,bera2018return,detomasi2019survival,tikhonov2019statistics,biroli2020anomalous,detomasi2020subdiffusion} can be harnessed to understand the Hilbert-space return probability notwithstanding the crucial role of correlations on the Hilbert-space graph~\cite{roy2020fock,roy2020localisation}.
This in turn can provide important insights into the local autocorrelations.

It is worth mentioning that while probing real-space dynamical correlations has been experimentally possible for a few years now~\cite{schreiber2015observation,bordia2017periodically}, experimentally probing the dynamics directly on the Hilbert/Fock space has also become possible very recently~\cite{yao2022observation}. Our results can therefore explain the possible connections between the two complementary sets of experiments and how to understand the results of one in terms of the other.

As a concrete setting, we consider a chain of interacting quantum spins-1/2, which we denote by the set of Pauli matrices, $\{\sigma^\mu_i\}$ with $\mu=x,y,z$, and $i$ labels the real-space site. 
We focus on the \new{infinite-temperature spin}-autocorrelation,
\eq{
	\Ct(t) = \frac{1}{L}\sum_{i=1}^L\frac{1}{\nh}\mathrm{Tr}[\sigma^z_i(t)\sigma^z_i(0)]\,,
	\label{eq:Ct}
}
with $\nh$ the Hilbert-space dimension and $L$ the system size.
The autocorrelation in Eq.~\ref{eq:Ct} mandates a natural choice for the Hilbert-space basis, namely, $\sigma^z$-product states, $\ket{\vec{s}_I} = \ket{s_{I,1},s_{I,2}\cdots,s_{
I,L}}$ with $s_{I,i}=\pm 1$.
The Hilbert-space return probability is defined as
\eq{
	\Rt(t) = \nh^{-1}\sideset{}{_{I=1}^{\nh}}\sum|\braket{\vec{s}_I|U(t)|\vec{s}_I}|^2\,,
	\label{eq:Rt}
}
where $U(t)$ is the time-evolution operator. 

Two crucial points are to be noted here.
First, due to the extensive connectivity of the Hilbert-space graph, $\Rt(t)$ is exponentially suppressed in $L$ for any non-zero $t$. 
As such, $\Rt(t)$ does not admit a well-defined thermodynamic limit and it is $[\Rt(t)]^{1/L}$ or equivalently, $\ln[\Rt(t)]/L$ that does.
This is a crucial distinction from return probabilities on hierarchical graphs with fixed $\mathcal{O}(1)$ connectivities~\cite{biroli2017delocalized,bera2018return} where $\Rt(t)$ itself is well-defined in the thermodynamic limit.
The second point is regarding the late-time saturation of the autocorrelation and the return probability. In the ergodic phase, $\Ct(t)$, decays in a universal fashion for arbitrarily long times in the thermodynamic limit, eventually saturating to 0 at infinite times. On the other hand, $[\Rt(t)]^{1/L}$ saturates to a finite value as $t\to\infty$ in the thermodynamic limit, irrespective of whether the dynamics is in an MBL or an ergodic phase. This can be understood via \eq{\lim_{t\to\infty}\Rt(t) = \nh^{-1}\sum_{I}\sum_{\omega}|\braket{\omega|\vec{s}_I}|^4\,,}
which is nothing but the average inverse participation ratio (IPR) of the eigenstates, $\{\ket{\omega}\}$, of the Hamiltonian or the Floquet unitary which generates the time evolution.
Since the IPR $\sim\nh^{-\tau}$ with $\tau=1$ in the ergodic phase and $0<\tau<1$ in the MBL phase~\cite{deluca2013ergodicity,mace2019multifractal,roy2021fockspace}, and $\nh\sim e^{\gamma L}$ with $\gamma>0$, we have $\mathcal{R}_\infty^{1/L}\equiv\lim_{t\to\infty}\Rt(t)^{1/L} \sim e^{-\gamma\tau}$, an $\mathcal{O}(1)$ constant. Moreover, since $[\Rt(t)]^{1/L}$ is independent of $L$, the timescales associated to this saturation are also independent of $L$.

An important outcome of this is that it is how the return probability saturates to its infinite-time value at $t\gg 1$ that encodes how the real-space autocorrelation decays at arbitrarily long times in the ergodic phase. Quantitatively, in this work, we find that
\eq{
\lim_{L\to\infty}\frac{1}{L}\ln\left[\frac{\Rt(t)}{\mathcal{R}_\infty}\right] \sim [\Ct(t)]^\alpha\,;\quad t\gg 1\,,
\label{eq:Rt-Ct-rel}
}
where $\alpha>0$ is a model- and parameter-dependent, non-universal constant. This constitutes the central mathematical result of this work.
Before delving into its derivation, let us discuss the implications of the relation in Eq.~\ref{eq:Rt-Ct-rel}. For $t\gg 1$, in the ergodic phase, we have $\Ct(t)\ll 1$ such that Eq.~\ref{eq:Rt-Ct-rel} can be recast as 
\eq{
[\Rt(t)]^{1/L}-[\Rt_\infty]^{1/L} \sim [\mathcal{R}_\infty]^{1/L} [\Ct(t)]^\alpha\,.
}
The physical consequence of this is that, for a model without any conservation laws, where the spin autocorrelation may decay as a stretched exponential~\cite{gopalakrishnan2016griffiths,lezama2019apparent}, $\Ct(t)\sim \exp[-(t/\tau)^\beta]$ with $0<\beta\le 1$, the return probability also approaches its saturation value as a stretched exponential,
\eq{
[\Rt(t)]^{1/L}-[\Rt_\infty]^{1/L} \sim [\mathcal{R}_\infty]^{1/L} \exp[-(t/\tau^\prime)^\beta]\,,
}
albeit with a rescaled timescale $\tau^\prime\equiv \tau/\alpha^{1/\beta}$ but with same same stretching exponent $\beta$. On the other hand, for a model with conserved total $\sigma^z$-spin, the autocorrelation decays (sub)diffusively~\cite{gopalakrishnan2016griffiths,luitz2016anomalous}, $\Ct(t)\sim t^{-1/z}$ with $z\ge 2$ such that the return probability also approaches it saturation as a power-law in time, but with a renormalised power-law exponent $\alpha/z$,
\eq{
[\Rt(t)]^{1/L}-[\Rt_\infty]^{1/L} \sim [\mathcal{R}_\infty]^{1/L} t^{-\alpha/z}\,.
}
The upshot of the above is that a stretched-exponential or a power-law decay of the autocorrelation in time manifests itself into an analogous stretched-exponential or power-law approach of the return probability to its saturation value respectively; this constitutes the central physical implication of the result in Eq.~\ref{eq:Rt-Ct-rel}. In this way, the return probability unambiguously encodes the universal dynamics of the local spin autocorrelations and distinguishes between the absence or presence of a corresponding conservation law.

Having stated the central results of the work, we next turn towards their derivations. The relation between the autocorrelation and the return probability proceeds via two steps,
\begin{itemize}
    \item[(i)] by first relating the autocorrelation to the spread of the time-evolving wavefunction on the Hilbert-space graph,
    \item[(ii)] and then using the conservation of total probability on Hilbert space to relate the spread to the return probability.
\end{itemize} 
The first step above closely follows the analysis in Ref.~\cite{roy2021fockspace,creed2023probability}.
In order to quantify the spread of the state on the Hilbert space, we need a measure of distance on the Hilbert-space graph. The Hamming distance, namely the number of spins which are anti-aligned between two configurations is a natural choice. Formally defined as 
\eq{
r_{IJ} \equiv \frac{1}{4}\sum_{i=1}^L(s_{I,i}-s_{J,i})^2 = \frac{L}{2}-\frac{1}{2}\sum_{i=1}^Ls_{I,i}s_{J,i}\,,
\label{eq:ham-dist}
}
it constitutes a first link between Hilbert-space quantities and local spin observables~\cite{detomasi2020rare,roy2021fockspace,creed2023probability}. With the notion of distance on the Hilbert-space graph so defined, the spread of the wavefunction therein can be quantified as
\eq{
\mathcal{G}(r,t)=\frac{1}{\nh}\sum_{\substack{I,J:\\r_{IJ}=r}} |\braket{\vec{s}_J|U(t)|\vec{s}_I}|^2\equiv \frac{1}{\nh}\sum_{\substack{I,J:\\r_{IJ}=r}} \mathcal{G}_{IJ}(t)\,.
\label{eq:G-def}
}
Physically, $\mathcal{G}(r,t)$ can be interpreted as, starting from a spin-configuration (a node on the Hilbert-space graph), it is the total probability density of the state on nodes at distance $r$ from the initial node on the Hilbert-space graph. Conservation of probability naturally implies $\sum_{r=0}^L\mathcal{G}(r,t)=1$. As such $\mathcal{G}(r,t)$ is a {\it bona fide} probability distribution on $r$ with moments 
\eq{
\braket{r^n(t)} = \sum_{r=0}^L r^n\mathcal{G}(r,t) = \frac{1}{\nh}\sum_{I,J} r_{IJ}^n\mathcal{G}_{IJ}(t)\,.
\label{eq:r-moments}
}
The notion of distance defined in Eq.~\ref{eq:ham-dist} is particularly useful and relevant as the autocorrelation in Eq.~\ref{eq:Ct} can be recast as
\eq{
\Ct(t) &= \frac{1}{\nh}\sum_{I,J}|\braket{\vec{s}_J|U(t)|\vec{s}_I}|^2\frac{1}{L}\sum_i s_{I,i}s_{J,i}\,\nonumber\\
&=1-\frac{1}{\nh}\sum_{I,J}\mathcal{G}_{IJ}(t)\frac{2r_{IJ}}{L}=1-2\frac{\braket{r(t)}}{L}\,,\label{eq:Ct-Gt}
}
{\new{where in the second line we have used the definitions of $r_{IJ}$ and $\mathcal{G}_{IJ}(t)$ in Eqs.~\ref{eq:ham-dist} and \ref{eq:G-def} as well as that of $\braket{r(t)}$ in Eq.~\ref{eq:r-moments}.}}
% Using Eq.~\ref{eq:r-moments} can be expressed as 
% \eq{
% \Ct(t) = 1-2\braket{r(t)}/L\,.
% \label{eq:Ct-Gt}
% }
In this way, the spread of the wavefunction on the Hilbert-space graph is directly related to the autocorrelation which constitutes the first of the two aforementioned steps in the derivation. At the same time, note that
\eq{
\Rt(t) = \mathcal{G}(r=0,t)\,.
\label{eq:Rt-Gt}
}
Since the total $\mathcal{G}(r,t)$ is conserved, its decay at $r=0$ which encodes the return probability, is intimately connected to how it spreads which in turn carries information of the autocorrelation. That the dynamics of the autocorrelation and the return probability are intimately related to each other is thus put on a formal footing via Eq.~\ref{eq:Ct-Gt} and Eq.~\ref{eq:Rt-Gt} with the common thread being the Hilbert-spatiotemporal spread of the wavefunction characterised by $\mathcal{G}(r,t)$. To establish a precise relation between the autocorrelation and the return probability, we therefore need to understand the profile of $\mathcal{G}(r,t)$.
% which we turn to next.

\begin{figure}
\includegraphics[width=\linewidth]{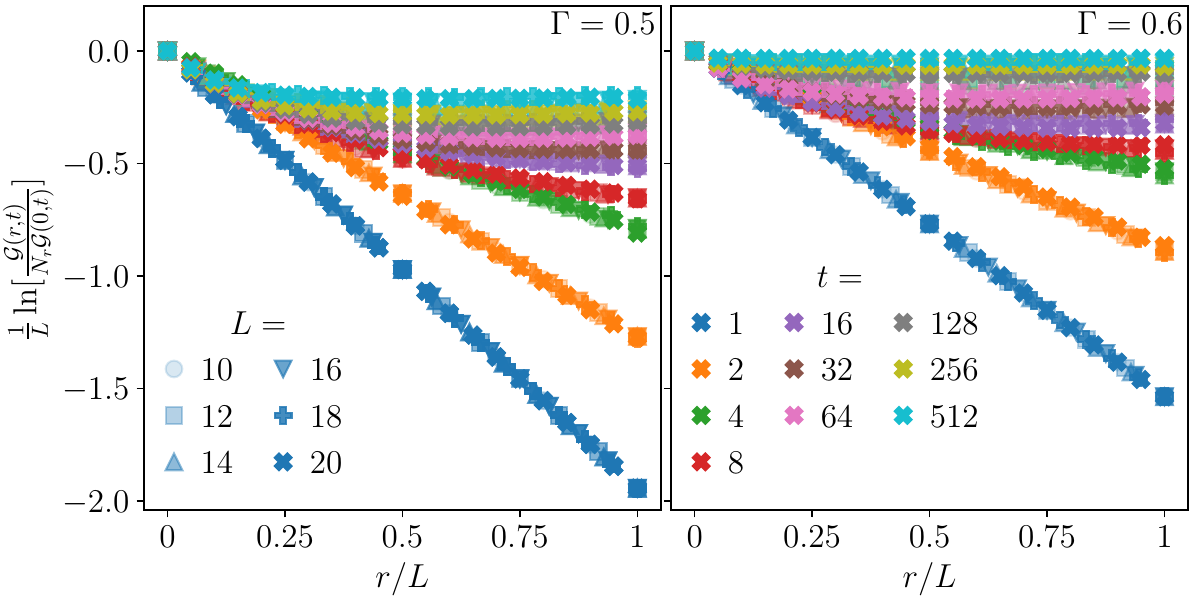}
\caption{The quantity $L^{-1}\ln[\overline{\mathcal{G}(r,t)}/N_r\overline{\mathcal{G}(0,t)}]$ as a function of $r/L$ for different $L$ and $t$, numerically computed for the disordered Floquet spin chain described in Eq.~\ref{eq:UF} and Eq.~\ref{eq:HXHZ}. Different intesities/markers correspond to different $L$ and the colours to different $t$ as noted in the legends. The two panels correspond to different parameter values, both in the ergodic phase. The collapse of the data for different $L$ implies the large-deviation form in Eq.~\ref{eq:large-dev}.}
\label{fig:Grt-large-dev}
\end{figure}

\new{Since $\mathcal{G}(r,t)$ is the sum of the probability densities at time $t$ over all nodes at distance $r$ from the initial time, upon disorder averaging\footnote{\new{While the disorder averaging is natural for a disordered system, our theory works equally well for a clean ergodic system~\cite{supp} where the average over initial states, implicit in the definitions in Eq.~\ref{eq:Ct} and Eq.~\ref{eq:Rt} is sufficient.}}, (denoted by an overline henceforth) it can be expressed as}
% Upon disorder averaging (denoted by an overline henceforth), we can express
\eq{
\overline{\mathcal{G}(r,t)} = N_r \mathcal{F}(r,t,L)\,,
\label{eq:Grt-ansatz}
}
where $N_r$ is the number of Hilbert-space nodes at distance $r$ from any initial node, and $\mathcal{F}(r,t,L)$ \new{has the physical meaning of the average probability density on any one of them at time $t$.} Note that for $L\gg 1 $, $N_r$ has a Gaussian profile,
\eq{
N_r \approx \nh \frac{e^{-\frac{(r-L/2)^2}{L/2}}}{\sqrt{\pi L/2}}\,,
\label{eq:Nr-gaussian}
}
such that it is extremely sharply peaked at $r=L/2$ \new{as the width scales only as $\sqrt{L}$}.
At the same time, for any $t>0$, $0<\overline{\Ct(t)}<1$ with the inequalities being strict implying that $\overline{\braket{r(t)}}\sim L$. These two aspects together strongly hint (albeit not conclusively) that $\mathcal{F}(r,t,L)$ has a large deviation form
\eq{
\overline{\mathcal{G}(r,t)} = N_r e^{-LF(x,t)}=  N_re^{-L[f(x,t)+g(t)]}\,,
\label{eq:large-dev}
}
where $x=r/L$ \new{and $F(x,t)$ can be physically interpreted as the rate function associated to the large-deviation form of $\mathcal{F}(r,t,L)$. We decompose the function $F(x,t)=f(x,t)+g(t)$ with $f(x=0,t)=0$ purely for convenience as it allows us to separate out the $r=0$ contribution into $g(t)$ and } 
% We split the function $F(x,t)$ without loss of generality such that $f(0,t)=0$ and $F(0,t)=g(t)$. This is purely for convenience as it allows us to 
express the return probability, using Eq.~\ref{eq:Rt-Gt}, as 
\eq{
\overline{\Rt(t)} = e^{-L g(t)}\,.
\label{eq:Rt-gt}
}
The large-deviation form of $\overline{\mathcal{G}(r,t)}$ is indeed borne out conclusively by exact numerical results (see Fig.~\ref{fig:Grt-large-dev}) on a disordered spin-1/2 chains, which we discuss later.

The key point here is that the large-deviation form of $\overline{\mathcal{G}(r,t)}$ implies that the moments, defined in Eq.~\ref{eq:r-moments}, are governed by the saddle points in the corresponding integrals.
The saddle point, which we denote as $x_\ast(t)$, is given by the solution to the equation
\eq{
	\partial_xf(x,t)|_{x=x_\ast(t)} +4(x_\ast(t)-1/2)=0\,,
	\label{eq:sp}
}
where the second term comes from the argument of the exponential in Eq.~\ref{eq:Nr-gaussian}.

The zeroth moment, due to the the normalisation of $\mathcal{G}(r,t)$, is equal to unity, which yields a relation for $g(t)$ in terms of $f(x_\ast(t),t)$ and $x_\ast(t)$, such that return probability, using Eq.~\ref{eq:Rt-gt}, can be expressed as 
\eq{
\overline{\Rt(t)}^{1/L} \overset{L\to\infty}{\approx} \nh^{-1/L}e^{f(x_\ast(t),t)+2[x_\ast(t)-1/2]^2}\,.
\label{eq:Rt-sp}
}
In the ergodic phase, since $\overline{\mathcal{R}_\infty}\sim \nh^{-1}$, the above equation yields
\eq{
\lim_{L\to\infty}\frac{1}{L}\ln\left[\frac{\overline{\Rt(t)}}{\overline{\mathcal{R}_\infty}}\right]=f(x_\ast(t),t)+2[x_\ast(t)-1/2]^2\,,
\label{eq:Rt-Rinf-sp}
}
which is an explicit relation between the return probability and, the saddle point $x_\ast(t)$ and the function $f(x,t)$ at the saddle point. In the same spirit, the first moment of $\overline{\mathcal{G}(r,t)}$ along with Eq.~\ref{eq:Ct-Gt} leads to 
\eq{
\overline{\mathcal{C}(t)} = 1-2x_\ast(t) = \frac{1}{2}\partial_xf(x,t)|_{x=x_\ast(t)}\,,
\label{eq:Ct-sp}
}
which constitutes a precise relation between the autocorrelation and the saddle points or equivalently, the behaviour of the function $f(x,t)$ around the saddle point. In the second equality above, we have used the equation for the saddle point, \eqref{eq:sp}.

\begin{figure}
\includegraphics[width=\linewidth]{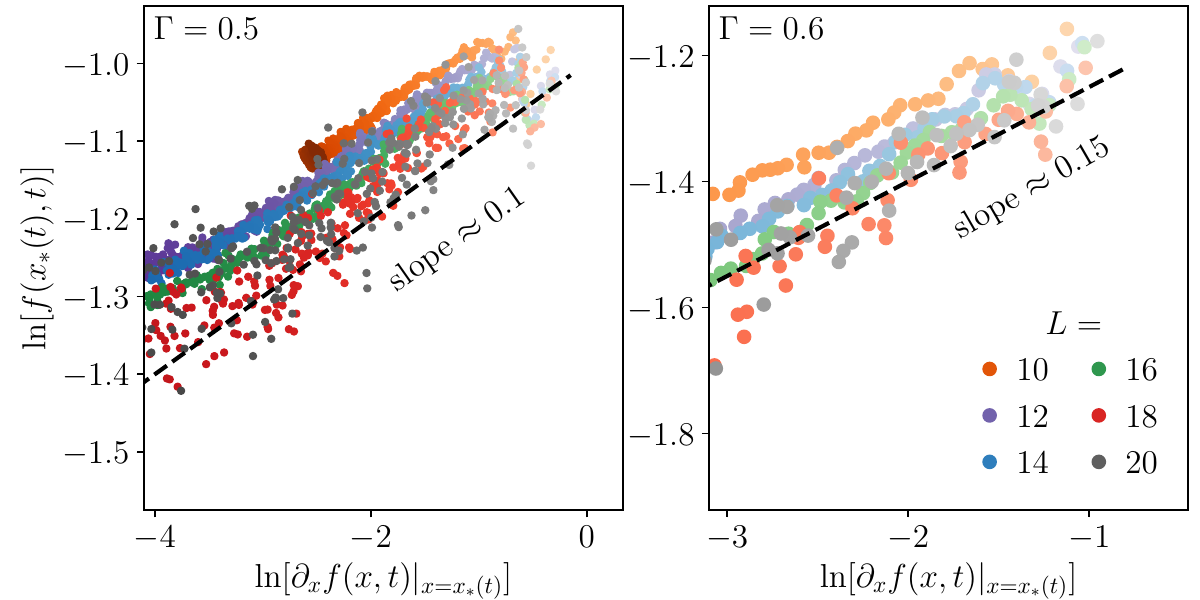}
\caption{Logarithms of the functions $f(x_\ast(t),t)$ and $\partial_xf(x,t)|_{x_\ast(t)}$, corresponding to the plots in Fig.~\ref{fig:Grt-large-dev}, shown as scatter plots where $x_\ast(t)$ denotes the saddle points given by Eq.~\ref{eq:sp}. Different colours correspond to different $L$ and different intensities denote data points for different times in the range $t\in [1,1000]$ with darker colours denoting later times. The linear behaviour supports the conjecture in Eq.~\ref{eq:fx-dfx}.}
\label{fig:fx-dfx}
\end{figure}

From Eq.~\ref{eq:Rt-Rinf-sp} and Eq.~\ref{eq:Ct-sp}, it is clear that it is the saddle point and the behaviour of the function $f(x,t)$ around it that forms the bridge between the approach of the return probability to its saturation value and the autocorrelation. In particular, the second term in the right-hand side of Eq.~\ref{eq:Rt-Rinf-sp} can be identified as $[\overline{\Ct(t)}]^2/2$. 
However, the last remaining puzzle is to understand how is the information of $\overline{\Ct(t)}$ contained in $f(x_\ast(t),t)$. The key point here is that as $t\to\infty$, both $f(x_\ast(t),t)$ and $\partial_xf(x,t)|_{x_\ast(t)}$ tend towards 0 in the ergodic phase. Both of these can be understood from the fact that at very late times, the wavefunction on average spreads out homogeneously over the entire Hilbert-space graph. As such, $\mathcal{F}(r,t,L)$ in Eq.~\ref{eq:Grt-ansatz} becomes independent of $r$, which in turn implies that $f(x,t)\to 0$ as $t\to\infty$. This also means that the only dependence on $r$ in $\overline{\mathcal{G}(r,t)}$ comes from $N_r$ which, recall, is sharply peaked at $r=L/2$. This implies $x_\ast(t\to\infty)\to 1/2$ which in turn, via Eq.~\ref{eq:sp}, leads us to conclude that $\partial_xf(x,t)|_{x_\ast(t)}\to 0$ as $t\to\infty$. As both $f(x_\ast(t),t)$ and $\partial_xf(x,t)|_{x_\ast(t)}$ tend to 0 as $t\to\infty$, a natural conjecture for $t\gg 1$ is
\eq{
f(x_\ast(t),t)\sim \left[\partial_x f(x,t)|_{x=x_\ast(t)}\right]^{\alpha^\prime}\sim [\overline{\Ct(t)}]^{\alpha^\prime}\,,
\label{eq:fx-dfx}
}
where $\alpha^\prime$ is a non-universal constant. Numerical results, shown in Fig.~\ref{fig:fx-dfx}, albeit severely constrained by system sizes, provide strong evidence in support of the conjecture. Using Eq.~\ref{eq:fx-dfx} in Eq.~\ref{eq:Rt-Rinf-sp}, we obtain
\eq{
\lim_{L\to\infty}\frac{1}{L}\ln\left[\frac{\overline{\Rt(t)}}{\overline{\mathcal{R}_\infty}}\right]\approx a[\overline{\Ct(t)}]^{\alpha^\prime}+\frac{1}{2}[\overline{\Ct(t)}]^2\sim [\overline{\Ct(t)}]^{\alpha}\,,
\label{eq:final-res}
}
where $\alpha = \mathrm{min}(\alpha^\prime,2)$. This concludes the analytical derivation of the main result of the work stated in Eq.~\ref{eq:Rt-Ct-rel}.

We next discuss the numerical results, which informed the analytical derivation above. As a concrete model, we employ a disordered, Floquet (periodically driven) Ising spin-1/2 chain without any explicitly conserved quantities~\cite{zhang2016floquet}. The time-evolution operator over one period, dubbed as the Floquet unitary, is given by 
\eq{
U_F = \exp[-i\tau H_X]\,\exp[-i\tau H_Z]\,,
\label{eq:UF}
}
with 
\begin{equation}
    \begin{split}
        H_X &=g\Gamma\sum_{i=1}^L \sigma^x_i\\
        H_Z &= \sum_{i=1}^L [\sigma^z_i\sigma^z_{i+1} + (h+g\sqrt{1-\Gamma^2}\epsilon_i)\sigma^z_i]
    \end{split}\,,
    \label{eq:HXHZ}
\end{equation}
where $\sigma_i$'s are Pauli matrices representing the spins-1/2 and $\epsilon_i\sim \mathcal{N}(0,1)$ are standard Normal random numbers. Following Ref.~\cite{zhang2016floquet} we take $g = 0.9045$, $h = 0.809$, and $\tau = 0.8$. For these parameters, there is a putative many-body localisation transition at $\Gamma_c\approx 0.3$ with the model in an ergodic phase for $\Gamma>\Gamma_c$ and in an MBL phase for $\Gamma<\Gamma_c$. It is the former, on which we mostly focus.

In Fig.~\ref{fig:Grt-large-dev}, we plot $L^{-1}\ln[\overline{\mathcal{G}(r,t)}/N_r\overline{\mathcal{G}(0,t)}]$ as a function of $r/L$ for different $t$ and $L$. From Eq.~\ref{eq:large-dev} we expect this quantity to be $f(x,t)$. We find an excellent collapse of the data for different $L$ which conclusively shows that $\overline{\mathcal{G}(r,t)}$ has a large-deviation form and the argument of the exponential in Eq.~\ref{eq:large-dev} is function of $x=r/L$. Also, note from Fig.~\ref{fig:Grt-large-dev} that as $t$ increases, $f(x,t)$ becomes more and more flat at a smaller and smaller value. This lends support to the analytical understanding discussed earlier that both $f(x_\ast(t),t)$ and $\partial_xf(x,t)|_{x_\ast(t)}$ tend towards 0 as $t\to\infty$. The question then is how they vanish relative to each other -- a conjecture for this is provided in Eq.~\ref{eq:fx-dfx}.
In Fig.~\ref{fig:fx-dfx}, we plot $\ln[\partial_xf(x,t)|_{x_\ast(t)}]$ against the corresponding $\ln[f(x_\ast(t),t)]$ for different $t$ and $L$. The linear behaviour shows that the conjecture is indeed consistent, particularly at $t\gg 1$. In order to extract the saddle points numerically, we interpolate the functions $f(x,t)$ (shown in Fig.~\ref{fig:Grt-large-dev}), and have checked the results are robust to interpolation orders.

For completeness, we mention that we also performed numerical simulations on a disordered, Floquet XXZ chain~\cite{ponte2015many} wherein $H_X = \Gamma\sum_{i}[\sigma^x_{i}\sigma^x_{i+1}+\sigma^y_{i}\sigma^y_{i+1}]$. This model possesses total-$\sigma^z$ conservation unlike the Ising model in Eqs.~\ref{eq:UF}-\ref{eq:HXHZ}. The numerical results for the Floquet XXZ model are qualitatively similar to those of the Ising model, and hence we omit them to avoid repetition. This concludes our discussion of exact numerical results, which provides conclusive evidence for the validity of the analytic calculations presented earlier based on the saddle points of the function $f(x,t)$.

\new{Before we close and summarise, we make some remarks about the situation MBL phases. Contrary to ergodic phases, $\mathcal{C}_\infty\equiv\mathcal{C}(t\to\infty)\to\mathcal{O}(1)$ in the MBL phase and hence Eq.~\ref{eq:Rt-Ct-rel} picks up corrections on the right hand side. On the Hilbert-space graph, the function $f_\infty(x) \equiv f(x,t\to\infty)$ saturates to a non-trivial function as opposed to $f_\infty(x)=0$ in the ergodic phase in such a way that the saddle point $x_\ast(t\to\infty)$ saturates to a value strictly less than 1/2, ensuring that $\mathcal{C}_\infty\to\mathcal{O}(1)$ (see Eq.~\ref{eq:Ct-sp}). While we relegate the derivation to the supplementary material~\cite{supp}, the result analogous to Eq.~\ref{eq:Rt-Ct-rel} in the MBL phase is
\eq{
\lim_{L\to\infty}\frac{1}{L}\ln\left[\frac{\Rt(t)}{\mathcal{R}_\infty}\right] \sim [\Ct(t)-\mathcal{C}_\infty]^2\,;\quad t\gg 1\,,
\label{eq:Rt-Ct-rel-MBL}
}
provided the power-law decay of $[\Ct(t)-\mathcal{C}_\infty]^2\sim t^{-2b}$, predicted in Ref.~\cite{serbyn2014quantum} is slow enough which in turn is controlled by $b\propto \xi$ and the smallness of the $\ell$-bit localisation length $\xi$.
We speculate that the exponent of 2 on the RHS appears universally in the MBL phase as the $f_\infty(x)$ admits a well-defined Taylor expansion around $x_\ast(t\to\infty)$ with the leading linear term being finite in general~\cite{supp}.
}

To summarise, we derived an explicit relation, stated in Eq.~\ref{eq:Rt-Ct-rel}, between the temporal decay of spin autocorrelation and the approach of the return probability to its saturation value at late times. The central point of the result is that the functional form of the autocorrelation decay is mirrored by the return probability's temporal approach to its saturation. In this way, it bridges explicitly two complementary but apparently disconnected aspects of the dynamics of disordered, interacting quantum many-body systems. 
This is potentially a significant advance as the return probability holds a lot of promise for analytic computations by exploiting the wealth of literature on single-particle dynamics on disordered, high-dimensional graphs. The results presented in this work can then, in turn, be directly used to gain analytical insights into the dynamics of real-space local observables. 
The exact numerical results presented in this work were restricted to corroborating the large-deviation form in Eq.~\ref{eq:large-dev} and verifying the conjecture in Eq.~\ref{eq:fx-dfx}, which provide compelling numerical support for the theory. It is nevertheless  desirable to have a direct numerical comparison of the main result in Eq.~\ref{eq:Rt-Ct-rel}. However, since this involves evolving systems with rather large Hilbert-space dimensions for very long times to access the approach of $\Rt(t)$ to its saturation, we keep it as a topic of a separate numerical work on its own right.

% {\color{red}\sout{In this work, we focussed on the spin autocorrelation. A natural question of immediate interest is to develop such a theory for the two-point spatiotemporal correlation, $\mathcal{C}(x,t)\equiv\braket{\sigma^z_{i+x}(t)\sigma^z_i}$, which can directly characterise anomalous transport of conserved quantities}}{~\cite{agarwal2015anomalous,luitz2016anomalous,bera2017density,roy2018anomalous}}. 
% {\color{red}\sout{The goal here would be relate the anomalous spatiotemporal profile of $\mathcal{C}(x,t)$ to an analogously anomalous decay of a suitable Hilbert-space dynamical correlation.}}

Looking further afield, it will be interesting to generalise the theory for the morphology of higher-point Hilbert-spatial dynamical correlations. This is with the motivation that such a theory can shed light into the anomalous dynamics of entanglement, both in the ergodic and in the MBL phases~\cite{bardarson2012unbounded,serbyn2013universal,lezama2019powerlaw} based on the mapping between bipartite entanglement and Hilbert-space correlations~\cite{roy2022hilbert}.

\begin{acknowledgments}
BP and SR thank S. Banerjee and D. E. Logan for useful discussions as well as Y. Bar Lev and A. Lazarides for useful comments on the manuscript. SR also thanks K. Damle, R. Sensarma, V. Tripathi, and members of DTP at TIFR Mumbai for raising interesting questions during an informal seminar which influenced parts of this work. 
The authors acknowledge support from the Department of Atomic Energy, Government of India, under project no. RTI4001. KK was supported by the Long-Term Visiting Students Programme at ICTS-TIFR. SR also acknowledges support from an ICTS-Simons Early Career Faculty Fellowship via a grant from the Simons Foundation (677895, R.G.). 
\end{acknowledgments}

\bibliography{refs}

\clearpage

\setcounter{equation}{0}
\setcounter{figure}{0}
\setcounter{page}{1}
\renewcommand{\theequation}{S\arabic{equation}}
\renewcommand{\thefigure}{S\arabic{figure}}
\renewcommand{\thesection}{S\arabic{section}}
\renewcommand{\thepage}{S\arabic{page}}

\onecolumngrid

\begin{center}
{\bf {\large{Supplementary Material: The connection between Hilbert-space return probability and real-space autocorrelations in quantum spin chains }}}\\
\medskip

Bikram Pain, Kritika Khanwal, and Sthitadhi Roy\\
{\it {\small International Centre for Theoretical Sciences, Tata Institute of Fundamental Research, Bengaluru 560089, India}}
\end{center}

\twocolumngrid
This supplementary material contains two sections. In Sec. I, we show that the central result of the work holds for an ergodic Hamiltonian without disorder whereas in Sec. II we describe how the results gets modified in the MBL phase.

\section*{I. Clean Ergodic Hamiltonian }
\begin{figure}[h]
\includegraphics[width=\linewidth]{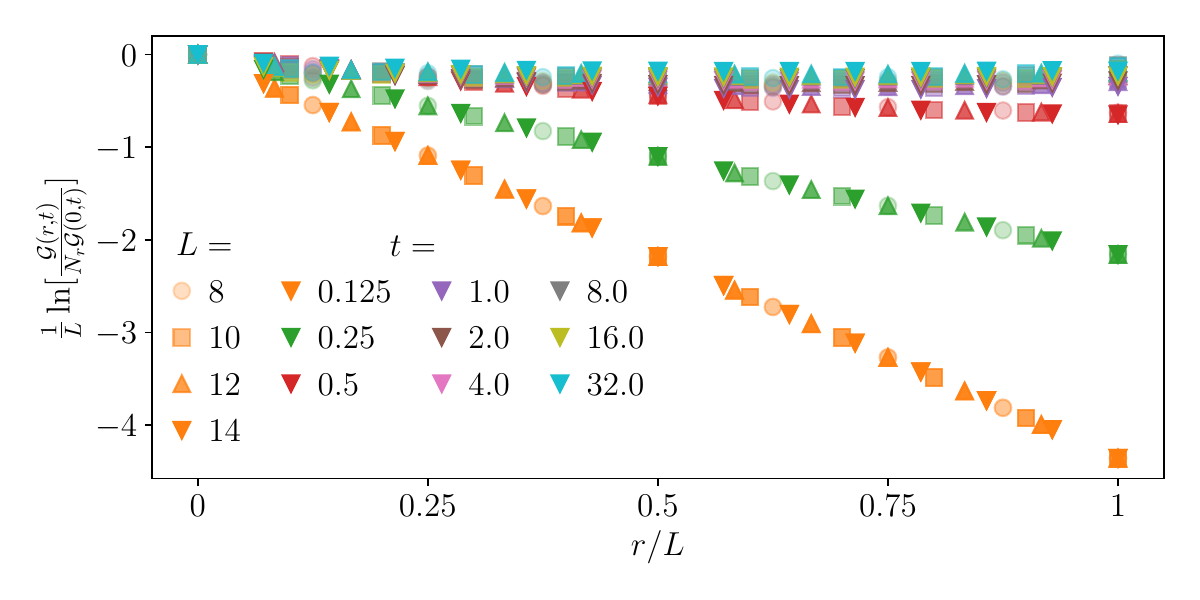}
\caption{Results analogous to those of Fig.~\ref{fig:Grt-large-dev} for the time-independent, clean Hamiltonian in Fig.~\ref{eq:Ham}.}
\label{fig:large-dev-ham}
\end{figure}

The central result of the work, Eq.~\ref{eq:Rt-Ct-rel}, is expected to be valid for any generic ergodic system. The numerical results in the main text supporting the central ingredient in the derivation of the result, namely the large deviation form in Eq.~\ref{eq:large-dev}, was shown for a Floquet system with moderate disorder \eqref{eq:UF}. Here we show that the results are qualitatively the same for an ergodic, time-independent Hamiltonian without disorder. This provides evidence for the validity of the result in clean, ergodic systems and also shows that a Floquet model is not necessary. For concreteness, we consider the Hamiltonian 
\eq{
H = \sum_{i=1}^L[\sigma^z_i\sigma^z_{i+1}+h\sigma^z_i+g\sigma^x_i]\,,
\label{eq:Ham}
}
with $h=0.809$ and $g=0.9045$. The results analogous to those in Fig.~\ref{fig:Grt-large-dev} are shown in Fig.~\ref{fig:large-dev-ham}. The collapse of the curves of $L^{-1}\ln[\mathcal{G}(r,t)/N_r\mathcal{G}(0,t)]$ for different $L$ at a fixed $t$ when plotted as a function of $x=r/L$ again provides conclusive evidence for the large-deviation form in Eq.~\ref{eq:large-dev}.

\section*{II. Many-body localised phase}

In this section, we discuss the fate of the results in an MBL phase. The result in Eq.~\ref{eq:Rt-Ct-rel} is valid only for ergodic systems as the left-hand side goes zero as $t\to\infty$ by definition whereas the right-hand side does so only for ergodic systems. In an MBL phase, on the other hand, $\mathcal{C}_\infty\equiv \mathcal{C}(t\to\infty)\to\mathcal{O}(1)$ value such that the right-hand side picks up corrections and the relation between the return probability and the autocorrelation gets modified to that in Eq.~\ref{eq:Rt-Ct-rel-MBL}. In the following, we outline its derivation.

\begin{figure}
\includegraphics[width=\linewidth]{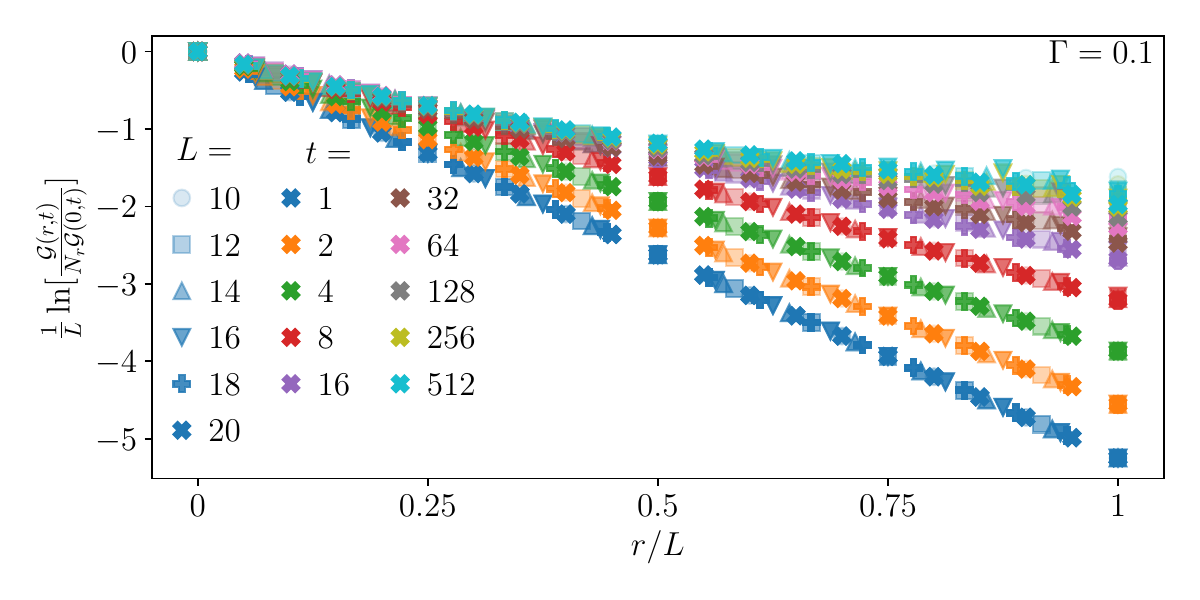}
\caption{Results analogous to those of Fig.~\ref{fig:Grt-large-dev} for the disordered Floquet model in Eq.~\ref{eq:UF} for $\Gamma =0.1 $ such that the model is in the MBL phase.}
\label{fig:large-dev-MBL}
\end{figure}

The large-deviation form of Eq.~\ref{eq:large-dev} continues to hold in this case as well, see Fig.~\ref{fig:large-dev-MBL}. However, contrary to ergodic systems where $f(x,t\to\infty)\to 0$ for all $x$, in an MBL phase, the rate function saturates to a nontrivial function of $x$ at infinite time which we will denote as $f_\infty(x) \equiv f(x,t\to\infty)$. This is also corroborated in the numerical results in Fig.~\ref{fig:large-dev-MBL}. Additionally, the saddle point saturates to value $x_\infty \equiv x_\ast(t\to\infty)$ which is strictly less than 1/2 which in turn ensures that $\mathcal{C}_\infty$ is a finite $\mathcal{O}(1)$ constant, see Eq.~\ref{eq:Ct-sp}. Moreover, the fractal scaling of the IPRs in the MBL phase implies $\overline{\mathcal{R}_\infty}^{1/L}\sim \nh^{-\tau/L}$ with $0<\tau<1$. The analogue of Eq.~\ref{eq:Rt-Rinf-sp} then takes the form
\begin{equation}
   \lim_{L\to\infty}\frac{1}{L}\ln\bigg[\frac{\overline{\mathcal{R}(t)}}{\overline{\mathcal{R}(\infty)}}\bigg]=(\tau-1)\ln2+f(x_*(t),t)+\frac{1}{2}\mathcal{C}^{2}(t)\,.
   \label{eq:rt/rinf-MBL}
\end{equation}
Note that as $t\to\infty$ the left-hand side in the above equation vanishes by definition which naturally implies 
\eq{
f_\infty(x_\infty) = -(\tau-1)\ln 2 -\mathcal{C}_\infty^2/2\,.
\label{eq:finf-xinf-MBL}
}

Our interest, however, lies in the limit of $t\gg 1$ and the right-hand side of Eq.~\ref{eq:rt/rinf-MBL} as $t$ approaches infinity. This entails understanding the corrections to the the function $f(x,t)$ relative to $f_\infty(x_\infty)$. While $f(x,t)$ can be formally expanded in a Taylor series around $x_\infty$, we cannot do so for $t$ as $t\gg 1$. Hence, we define $z\equiv 1/t$ and $\tilde{f}(x,z) \equiv f(x,t)$ and expand the latter around $(x=x_\infty,z=0)$ as 
\eq{
f(x_\ast(t),t) = &f_\infty(x_\infty) + \partial_{x}\tilde{f}(x,z)|_{\substack{x=x_\infty \\ z=0}}(x_\ast(t)-x_\infty)\nonumber\\ &+ \partial_{z}\tilde{f}(x,z)|_{\substack{x=x_\infty \\ z=0}}z + \cdots\,,
}
which, using the solutions to the saddle point equation in Eq.~\ref{eq:sp} and the relation between $\Ct$ and the saddle point in Eq.~\ref{eq:Ct-sp}, can be expressed as 
\eq{
f(x_\ast(t),t) \simeq &f_\infty(x_\infty) + \mathcal{C}_\infty(\mathcal{C}_\infty-\Ct) + \frac{B}{t}\,,
\label{eq:fxt-expand-MBL}
}
where $B=\partial_{z}\tilde{f}(x,z)|_{\substack{x=x_\infty \\ z=0}}$ is an $\mathcal{O}(1)$ constant. Using Eq.~\ref{eq:finf-xinf-MBL} and \ref{eq:fxt-expand-MBL} in Eq.~\ref{eq:rt/rinf-MBL}, we finally obtain
\eq{
\lim_{L\to\infty}\frac{1}{L}\ln\bigg[\frac{\overline{\mathcal{R}(t)}}{\overline{\mathcal{R}(\infty)}}\bigg]\simeq\frac{1}{2}(\mathcal{C}(t)-\mathcal{C}_{\infty})^{2} +\frac{B}{t}\,.
\label{eq:res-MBL}
}
From Ref.~\cite{serbyn2014quantum}, we expect the fluctuation of $\Ct(t)$ around $\mathcal{C}_\infty$ to decay to zero as $t^{-b}$ where $b\propto \xi$ with $\xi$ the characteristic $\ell$-bit localisation length in the MBL phase. Sufficiently deep inside the MBL phase, we expect $\xi\ll 1$ such that the second term on the right-hand side of Eq.~\ref{eq:res-MBL} can be neglected in favour of the first leading to
\eq{
\lim_{L\to\infty}\frac{1}{L}\ln\bigg[\frac{\overline{\mathcal{R}(t)}}{\overline{\mathcal{R}(\infty)}}\bigg]\sim(\mathcal{C}(t)-\mathcal{C}_{\infty})^{2} \,,
}
which is the result in Eq.~\ref{eq:Rt-Ct-rel-MBL}.

\end{document}